\documentclass[a4paper, 12pt]{article}
\usepackage[backend=biber]{biblatex}
\usepackage{amsmath,amssymb,amsfonts,amsthm,amscd,makeidx,stackrel,stmaryrd,float,epsfig}
\usepackage[left=2.5 cm,top=2.5 cm,bottom=2.5 cm,right=2.5 cm]{geometry}
\usepackage{amsthm}
\usepackage[english]{babel}
\usepackage[T1]{fontenc}
\usepackage[utf8]{inputenc}
\usepackage{csquotes}
\usepackage{array}
\usepackage{bm}
\usepackage{eucal}
\numberwithin{equation}{section}
\usepackage{graphicx}
\usepackage{subcaption}
\usepackage{here}
\usepackage{float}
\usepackage{url}
\usepackage{colortbl}
\usepackage[export]{adjustbox}
 \usepackage{siunitx}
\usepackage{booktabs,amsmath}
\usepackage{wrapfig}
\usepackage{lipsum}
\usepackage{textcomp}
\usepackage{gensymb}
\usepackage{siunitx}
\usepackage{tikz}
\usetikzlibrary{shapes,arrows}
\usepackage{verbatim}
\usepackage[colorlinks = true,
           linkcolor = blue,
           urlcolor  = blue,
            citecolor =red]{hyperref}

\title{\textbf{Analytic inversion of closed form solutions of the satellite's $J_2$ problem}}
\author{ Alessio Bocci, Giovanni Mingari Scarpello}

\date{}

\begin{document}
\maketitle
\begin{abstract}
This report provides some closed form solutions -and their inversion- to a satellite's bounded motion on the equatorial plane of a spheroidal attractor (planet) considering the $J_{2}$ spherical zonal harmonic.
The equatorial track  of satellite motion- assuming the co-latitude $\varphi$ fixed at $\pi/2$- is investigated: the relevant time laws and trajectories are evaluated  as combinations of elliptic integrals of first, second, third kind and Jacobi elliptic functions.  The new feature of this report is: from the inverse $t = t(c)$ to get the period $T$ of some functions $c(t)$ of mechanical interest and then to construct the relevant $c(t)$ expansion in Fourier series, in such a way performing the inversion. Such approach-which led to new formulations for time laws of a $J_{2}$ problem- is benchmarked by applying it to the basic case of keplerian motion, finding again the classic results through our different analytic path. 

\end{abstract}





\section{Introduction}
The so called $J_2$ problem -or main problem of satellite theory-  can be considered as one of the most important of Celestial Mechanics.
The attractions of bodies were first investigated by Newton \footnote{ See sections XII and XIII of the first book of \textit{Principia},1687} and derived by him on the basis of \textit{synthetic} processes.

All the motions of a particle are treated by him under the assumption the Earth is a perfect sphere, but a real fact of nature is that all rotating bodies in the universe are spheroidal.

The Earth has a rather slight equatorial bulge: its diameter is about 43 km wider at the equator than pole-to-pole distance, a difference close to 1/300 of the diameter\footnote{An Earth scaled down to a globe with equatorial diameter of 1 meter would have that difference of only 3 millimeters.}.
The spheroidal geometry has consequences on the motion of a particle:

\begin{itemize}
\item the potential cannot be calculated any more as though the masses were concentrated at its centre,  
\item the real motion is not planar.  
\end{itemize}

The orbit of a particle about an oblate spheroid is in general not closed geometrically, and then not periodic. But considering the orbits projected in a plane some classes of closed orbits can be found in which the motion can be periodic.

\section{Aim of this paper}

We hereby prepare the way for solving the particle ODEs of motion induced by  a spheroidal attractor.
We will recall the ODEs system of a bead in spherical coordinates: radius $\rho$, co-latitude $\varphi$ and azimuth $\theta$. The oblate spheroid's potential will be presented in its formulation till to the harmonic $J_2$ corrective term.\footnote{Such a problem is known as $J_2$ problem, the symbol $J$ comes from Harold Jeffreys, geodesist, (1891-1989).}.
We will consider the projection of the 3D-motion on the spheroid equatorial plane: o.s. we study a particular planar motion ($\varphi = \pi / 2$) modeled by two unknown functions $\rho(t)$ and $ \theta(t) $.

\section{Nomenclature}


The main symbols recurring through this paper are listed below:

\vspace{5mm}

\begin{tabular}{ l l }

  $A,\, B,\, C $ & spheroid principal central moments of inertia \\
   ${\rm cn},\, {\rm dn},\, {\rm sn}$ & Jacobi elliptic functions \\
  ${\rm am}(F|k)$ & amplitude of $F$ with modulus $k$ \\
  $b$& semi-minor axis of the meridian ellipse of spheroid \\
  $c$& light velocity in vacuo \\
  $e=\sqrt{\frac{R^2-b^2}{R^2}}$ & meridian ellipse eccentricity \\
  ${\bf e}_r$ & unity vector along the radial direction\\
  ${\bf e}_\theta$ & unity vector along the transversal direction\\
  ${\bf e}_\varphi$ & unity vector along the third direction\\
  $E_0$ &  satellite's total energy \\
    $E(\varphi|k)$ & incomplete elliptic integral of second kind \\
    &  of modulus $k$ and amplitude $\varphi$\\
    $F(\varphi|k)$ & incomplete elliptic integral of first kind \\
    & of modulus $k$ and amplitude $\varphi$\\
  $G$ &  universal gravitational constant \\
  $I $ & spheroid moment of inertia with respect to a mobile axis \\
   $J_n$ &  n.$^{\rm th}$ zonal spherical harmonic \\
  $J_2=\frac{C-A}{MR^2}$ & quadrupole term in the spheroidal potential \\
  $k_0$ & constant areal velocity \\
  & (= satellite angular momentum per unity mass) \\
  ${\bf K}(k)$ & complete elliptic integral of first kind of modulus $k$\\
  $M$ &  spheroidal mass \\
  $ODE$ & Ordinary Differential Equation\\
  $PDE$ & Partial Differential Equation\\
   $R$ & equatorial radius of the oblate spheroid \\
   ${\rm sn}(\varphi|k), {\rm cn}(\varphi|k),{\rm dn}(\varphi|k)$ &  Jacobi elliptic functions of amplitude $\varphi$ and modulus $k$  \\
  $t$ & time\\
  $T$ & motion period \\
  $\hat{T}$ & angular period \\
  $U(r)$ & gravity potential function \\
 $------$\\
 $\gamma$ & mass density function \\
$\nabla^2 $ & Laplace operator \\
$\varepsilon=\frac{R-b}{R}$ & spheroidal flattening or oblateness \\
$\eta$ & first perturbation coefficient \\
$\theta(t)$ & satellite instantaneous azimuth \\
$\mu$ & attractor gravitational coefficient $=GM$\\
$\nu$ & attractor volumic mass \\
$\Pi(\varphi, \alpha|k) $ & incomplete elliptic integral of third kind \\
& of amplitude $\varphi$, parameter $\alpha$, modulus $k$\\
$\varphi$ & satellite instantaneous co-latitude \\
$\rho (t) $ & instantaneous value of satellite polar radius  \\
$\rho_0 $ & satellite initial value of  radius \\
$\rho_1, \rho_2, \rho_3$ & cubic's roots and reference values during the satellite motion  \\
$\Omega$ & Earth angular rotation speed \\
$\omega_0$ & satellite start-up angular rotation speed \\

\end{tabular}

\vspace{0.3cm}

Some other (many) symbols (like $a_{n}, b_{n}, \sigma_{n}, \zeta_{n}$, and so on) are used throughout the text as ease variables without a specific physical nor geometrical  meaning and do not need to be listed here.

\section{The spheroidal attractive potential and the \texorpdfstring{$J_2$}{TEXT} approximation}

 If the attractor is a point (or a sphere\footnote{A modern \textit{analytic} proof by  a triple integration that $U(r)=MG/r$ for an attracting uniform sphere can be read in \cite{teodorescu1}.}) we assume it (or its centre) as origin. If its mass is $M$, it will exert on a unitary mobile mass a centripetal force given by:
  $$
  \vec{F}=-\frac{GM}{\rho^3} \vec{\rho}
  $$
where the minus sign occurs being the radial reference $\vec{\rho}$ always oriented in centrifugal sense. It follows:
$$
U(r)=\int _\infty^r \vec{F}\cdot{\rm d}\vec{\rho}=-GM \int _\infty^r \frac{{\rm d}\rho}{\rho^2}= GM/r
$$
being zero the potential at infinity.
What above in order to avoid the confusion of signs often affecting this subject: the attractive potential is then positive.

The gravitational potential satisfies the Laplace PDE:
$$\nabla^2 U(x,y,z)=0$$ 

in the space region outside the mass generating the field in such a way. Starting from such a PDE in spherical coordinates, by separation of variables, one finds that the a.m. $U$ depends on azimuth through functions $P_n(\cos\theta)$ having a rotational symmetry around the $z$ axis i.e. \textit{spherical zonal harmonics} or Legendre polynomials. Of course there are  also radial and co-latitude harmonics (see e.g. \cite{vinti1}), so that we have three sets of indexed non dimensional coefficients i.e. spherical harmonics: \textit{zonal} $J_n$, \textit{tesseral} $C_{n,m}$ and \textit{sectorial} $S_{n,m}$. Therefore, solving the a.m. Laplace PDE, one will find $U(\rho,\theta,\varphi)$ as a double series of spherical harmonics which is known to be convergent. The expression currently presented is restricted to the zonal ones only.

Let be $\theta$ the azimuth of the satellite read over the equatorial plane of the spheroid  of radius $R$ and of semi-minor axis $b$.
MacCullagh died in the same year (1847) credited for his formula which is referred in 1849 by Allman at p. 387 of \cite{allman} and \cite{marzari2020ring} as 
$$
U=\frac{M}{r'}+\frac{1}{r'^3}(A+B+C -3I)
$$
where $A, B, C$ are the spheroid central moments of inertia and $I$ the moment about the straight line joining its centre to the particle which is far $r'$ from it.
But such a expression is not operative having $I$ to be evaluated with respect to a variable direction. Analyzing it, one meets again the spherical zonal harmonics $J_n$. For the Earth's potential, the $J_n$ nondimensional values depend on the mass distribution. As we know, \cite{puca} p. 45, $\left|J_2/J_3\right|\sim\left|J_3/J_4\right|\sim 10^{-3}$: therefore, limiting ourselves to the first approximation, we will cut the expansion \footnote{Hoots and France \cite{hoo} provide a semi-analytical solution for the motion of an artificial Earth satellite under the gravitation and the drag of a dynamic atmosphere. The attraction is taken into account including the zonal harmonics $J_2$, $J_3$ and $J_4$. They employ mean motion, eccentricity, inclination, argument of perigee, right ascension of ascending node and mean anomaly and apply a generalized method of averaging to obtain a transformation of variables which removes the dependence on the fast variable, mean anomaly. As far as we know, this is the only -even if not purely analytical- treatment including zonal harmonics beyond $J_2$. 
} not going beyond $J_2$. In such a way the motion of a main point, working on MacCullagh formula-after some long course here omitted- $U(\rho,\theta,\varphi)$  is found to be ruled (see e.g. \cite{puca}) by:

 \begin {equation}\label {Pot}
\boxed{U(\rho,\varphi)=\frac{GM}{\rho}+\eta\frac{GM}{\rho^3}(1-3\cos^2\varphi)+o\,(\rho/R)}
\end {equation}
where $\rho >0$ is the magnitude of the satellite position vector in an inertial planetocentric frame, $\varphi$ its co-latitude and its azimuth $\theta$ does not appear. The frame reference origin is assumed at the centre of spheroidal mass whose only motion is just given by the $\Omega$ rotation speed -around its North-South axis- responsible of flatness.

Equation \eqref{Pot} for $U$ does not hold the satellite's azimuth $\theta$ due to the perfect axial symmetry of the attracting body; but depends on radius $\rho$ and on the co-latitude $\varphi$ of the attracted body. Furthermore such a formulation is not reliable everywhere, but within the order of magnitude $\rho>>R$.
Mind that $\eta$ is the {\it first perturbation coefficient} measuring the a-sphericity of the spheroid and  given by 
$$\eta=\frac{1}{2}R^2\,J_2$$ where $ J_2$ is the not-dimensional {\it quadrupole moment} (aka {\it second  zonal spherical harmonic}) $J_2=\tfrac{C-A}{MR^2}$ being $A,B,C$ (again) the moments of inertia with respect to the central principal axes of inertia: 
$$A=B=\frac{1}{5}M (R^2+b^2)\, , C=\frac{2}{5}MR^2.$$ 
The non dimensional quantity $J_2$ has a different value for each heavenly body: for the Earth it is close to $1.082\times10^{-3}$.
 
The shape of the Earth would be spherical in absence of rotation: the equatorial bulge existence is due to the balance on the spheroid between its self-gravitation and the centrifugal force coming from rotation\footnote{If Earth were to rotate faster or slower, its shape would change and therefore so would do its gravitational potential.}, \cite{oliver}. Such a bulge produces the main  perturbations of keplerian potential. 
Formula \eqref{Pot} does provide the gravitational potential {\bf outside and quite far from the Earth surface} as formed by the keplerian plus the first spheroidal correction whose sign is the same of the first term and then providing a rise in attraction. The potential useful to analyze phenomena lying {\bf on the Earth surface}, e.g. tidal phenomena, is obtained by adding to the gravitational one that is coming from the centrifugal force, given by $\frac{\nu^2}{2}\Omega^2\cos^2\varphi$, being $\nu$ and $\Omega$ the Earth mass density and angular rotation speed.
Of course the potential due to the bulge also depends on $\Omega$; if in fact the latter e.g. increase, it would increase the flatness too and therefore the bulge's attractive effect.

\subsection{Is the \texorpdfstring{$J_2$}{TEXT} problem integrable ?}
 
 Not many systems of differential equations-after the two body problem- arising in Celestial Mechanics were found integrable. An example is the motion of a rigid body about a fixed point: if there are enough first integrals -capable to give a coordinate system on the set of solutions-then it is possible to reduce the original ODEs system to an equation solvable by an explicit integral, see e.g. \cite{Minrit} p. 6.
 
After two centuries of hard investigations only few solutions have been produced. At mid 19$^{\rm th}$ century, J. Liouville (1809-1882) provided a general framework \cite{lio}.

 The Liouville theorem states simply: the solution of the equations of motion of a \textit{Liouville integrable system} is obtained by quadrature. The proof can be read e.g. in \cite{camb} p.8.
 
And what about the motion induced by an axisymmetric spheroidal potential? In his treatise (1873)  I. Todhunter \cite{tod} describes the contents of about 300 papers or books of about 40 authors.

Celletti-Negrini in \cite{celletti} make reference to the hamiltonian where the potential is written with all the zonal harmonics $J_n$ and expound \textit{the $J_n$ problem} integrability as an open problem. 

On the contrary, the so called \textit{$J_2$ problem} is non-integrable at all, as stated numerically by other authors, see \cite{sim2}. In \cite{celletti} the non-integrability is investigated proving the existence of a region of chaotic motions because for such a hamiltonian the Lerman's \cite{lerman} chaoticity condition is met. Finally, let us quote that Irigoyen and Sim\'{o} proved, \cite{sim}, the analytical non-integrability of this problem using Ziglin theorem and the Yoshida criterion for homogeneous potentials.
Our -more simple- case is the \textit{$J_2$ equatorial problem} and will be solved by means of a composition of elliptic functions and inverted through a a Fourier series expansion.

Brouwer, \cite{brouwer} takes not two but three terms of the potential. The equatorial plane  fixed in space is used as the $xy$ plane of rectangular coordinates. He determines a particular solution which is the circular one, and afterwards looks for a general solution by studying the orbits near to it, dealing with the same analytical problem as  Hill in his lunar theory.

Janssens writes in \cite{jans} that orbits under an additional force term $\rho^{-4}$ can be expressed by elliptic functions quoting Whittaker, but his observation has to be rejected as Whittaker analyzes -at p. 81 of \cite{whit}- the  orbit integrability under a  single positional force as $\rho^{-n}$ and then \textit{not} a combination of them; furthermore the equations nonlinearity does not authorize such a conclusion, as we will see.

The motion -we are going to analyze- is of course a 3D one, so that the body moves in an orbital plane which is not fixed, but rotates about the Earth axis in the opposite direction to the satellite, see \cite{king} who solves analytically the orbit in vacuo by means of perturbative approach: an article of difficult reading of more than 40 symbols without a nomenclature list.

Omonile and alii in \cite{pelag} obtain the motion differential equations in oblate spheroidal coordinates.

In \cite{cong} the authors write the Lagrange equations for the satellite motion and -including the $J_2$ term-a first order perturbation solution is given in terms of standard orbital parameters.

The oblate spheroid describes the shape of axially spinning objects  ranging from the small Earth to gigantic galaxies. A recent article \cite{Anne} discusses formulae by Gauss and MacCullagh about the gravitational potential and derive new formulae -much more long and involved- for the gravitational potential, but lacking of a reliable motivation.

As it concerns the nature of trajectories, already in 1910 it was clear (e.g. see MacMillan, \cite{Mac}), the trajectory of a particle about an oblate spheroid is not in general closed, and then the motion is, in general, not periodic from a geometric point of view: in such case the word ''orbit'' should be avoided. The failure of these trajectories to close in space arises from the incommensurability of the period of rotation of the line of nodes with the period of motion in the revolving plane. When these periods happen to be commensurable, they will result closed in space (= orbits) and the motion is periodic.

Jeze \cite{jeze}, completely focused on the use of elliptic functions describes well the problem but does not pay enough room to trajectory and periodicity questions.

\section{ODEs system for the \texorpdfstring{$J_2$}{TEXT} problem}

\subsection{The equatorial assumption}
We now assume that the co-latitude of the motion is fixed at the value $ \varphi=\pi / 2 $: we study the planar motion in which the effective 3D-motion is projected onto the equatorial plane. In short, we are looking for the equatorial shadow of  the motion as projected by parallel light rays coming from the infinite and orthogonal to that plane. In such a way we get the ODE system of the $J_2$ equatorial problem {\it in vacuo}:

\begin{equation} \label{syst}
\begin{cases}
\ddot\rho-\rho \dot\theta^2 +\dfrac{\mu}{\rho^2}+3\eta\dfrac{\mu}{\rho^4}=0\\
\vspace{0.2 cm}
\dfrac{{\rm d}}{{\rm d}t}(\rho^2\,\dot \theta)=0
\end {cases}
\end{equation}

with the fixed, known and positive quantities $\eta$ and $\mu$. By the second of \eqref{syst}
we get: 
\begin{equation}\label{areal}
k_0=\rho^2\,\dot \theta=\rho_0^2 \omega_0\,\, ,\forall  t\geq 0
\end{equation}
which is the constant magnitude of the moment of momentum: i.e. the areal velocity.

 Notice that the Kepler first law (elliptic trajectory for bounded motion) and the third (square of periods of planets) are true \textit{only} in the specific keplerian case of inverse square force, and then not in such a field. On the contrary, the second one (constant areal velocity) keeps its validity also in the $J_2$  perturbed spheroidal field.
 However the differential equations decouple, i.e. each of them is expressible in terms of either $\rho$ or $\theta$. Let us focus first on the radial one.

\section{Inverse radial time law}

Plugging the constant of areas \eqref{areal} in the first of \eqref{syst}, we obtain the radial equation:
\begin{equation} \label{rad}
\ddot\rho-\rho \left(\dfrac{k_0^2}{\rho^4}\right) +\dfrac{\mu}{\rho^2}+3\eta\dfrac{\mu}{\rho^4}=0\\
\end{equation}

whose integrating factor is $\dot\rho$, so that \eqref{rad} becomes:
$$
\dot\rho \ddot\rho-k_0^2\,\dot\rho \rho^{-3}+\mu\dot\rho \rho^{-2}+3\mu \eta\dot\rho \rho^{-4}=0
$$
or:
$$
\dfrac{{\rm d}}{{\rm d}t}\left(\dfrac{\dot\rho^2}{2}+\dfrac{k_0^2}{2\rho^2}-\dfrac{\mu}{\rho}-\dfrac{\mu \eta}{\rho^3}\right)=0\\
$$

The quantity within brackets has the dimension of an energy for unity mobile mass $(J/kg)$ and then is the total (kinetic radial+kinetic transverse+potential keplerian+potential spheroidal) energy which keeps constant during time being no dissipation.
We put:

\begin{equation}\label {energy}
\dot\rho^2+\frac{k_0^2}{\rho^2}-2\frac{\mu}{\rho}-2\mu \frac{\eta }{\rho^3}=E_0, \quad \text{with} \quad E_0 \in \mathbb{R}
\end{equation}

where such a constant (=double total energy) is completely set by 5 numbers only: 3 initial conditions ($v_0, \omega_0,\rho_0$) and 2 process data $\mu,\eta$:
$$
E_0=v_0^2+\omega_0^2\,\rho_0^2-2\frac{\mu}{\rho_0}-2\mu \frac{\eta }{\rho_0^3}
$$
We consider only bounded \footnote{A recent solution for \textit{unbounded} motion in the equatorial plane induced by an oblate spheroid can be found in \cite{gurfil}.} motions, so that $E_0 < 0$. 

When in a bounded state (like a satellite in orbit around the Earth) the body's negative potential energy is in absolute value greater than its kinetic energy (positive), so that its \textit{total} energy is negative: this is said \textit{gravity well}. Then, if the object tries to escape from Earth, it would fly up, converting kinetic energy in potential  (i.e. slow down in the gravitational field). But since it has less kinetic than potential energy, it will eventually stop and fall down. 

On the contrary, if this difference is larger than zero (or equal), it will have enough energy to reach infinity, i.e. get away from Earth: it is not gravitationally bound.

Referring to bounded orbits, Goldstein's Classical Mechanics \cite{gold}  says, p.78 :
\begin{quote}
This does not necessarily mean that the orbits are closed. All that can be said is that they are bounded, contained between two circles of radii $\rho_1$ and $\rho_2$ with turning points always lying on the circles.
\end{quote}

Bounded doesn't automatically mean closed: means that there could be closed paths. On the other hand you can have closed, but not bounded trajectories.
  In order to avoid mistakes we hereinafter put 

$$0>E_0=-\left|E_0\right|, \forall  t\geq 0$$
In such a way, \eqref{energy} becomes:
\begin{equation}
{\rm d}t=\mp\frac{1}{\sqrt{\left|E_0\right|}}\frac{\rho^{3/2}{\rm d}\rho}{\sqrt{-\rho^3+\frac{2\mu}{\left|E_0\right|}\rho^2-\frac{k_0^2}{\left|E_0\right|}\rho+\frac{2\mu\eta}{\left |E_0\right|}}}
\label{TEMP}
\end{equation}
Let us go to compute the law time-radius.
The \eqref{TEMP}, can be written as:
\begin{equation}
{\rm d}t=\mp\frac{1}{\sqrt{\left|E_0\right|}}\frac{\rho^{3/2}{\rm d}\rho}{\sqrt{(\rho-\rho_1)(\rho-\rho_2)(\rho_3-\rho)}}
\label{rhointegral}
\end{equation}
Up to this point, before going on with integration, a short discussion on the cubic's roots $\rho_1<\rho_2<\rho_3$ is necessary. 
 
\begin{enumerate}
\item
Due to the practical values of data involved in the process, we computed\footnote{See e.g. later at the sample problem, sect.13} $\rho_1<R$, i.e. its value would fall inside the Earth and then is meaningless for us. As a consequence,
$\rho_1$ can be ignored. We basically then have 3 cases according to the mutual relationship between $\rho_2$ and $\rho_3$ only.
\item
If $\rho_2$ and $\rho_3$ are both real with $\rho_2 \neq\rho_3$, then the  admissible region of motion is just the open interval $(\rho_2,\rho_3)$ and the initial condition $\rho_0$ must be held in it. We will have a planar motion whose radial displacement starts from $\rho_0$ and oscillates-theoretically restless, due to the vacuum assumption- between $\rho_2$ and $\rho_3$.
\item
If $\rho_2=\rho_3$ are both real, then the quantity under square root in \eqref{rhointegral} can be written as $(\rho_1-\rho).D^2$ where D is a real constant.Then the admissible region of motion is $\rho_<\rho_1$ which is of course out of sense, due to the practical inconsistence of $\rho_1$.
\item

If the roots $\rho_2$ and $\rho_3$ are complex conjugate pair, then the quantity under square root in \eqref{rhointegral} can be written as $(\rho-\rho_1).C^2$ where C is another real constant, so that the radical reality simply requires $\rho>\rho_1$. In such a case time succeeds to be an ordinary integral and the body will do a progressive free motion, i.e. will not describe any revolution at all, but the mobile will escape far from the Earth. Such a case would then concern long-run explorations, and then is out of our purpose.
\end{enumerate}
Therefore we are going to take in account the scenario outlined in item 2).

Integrating \eqref{rhointegral}, we get:
\begin{equation}
t(\rho)=t_0 \pm \frac{1}{\sqrt{|E_0|}}[h(\rho)-h(\rho_0)]
\label{baba}
\end{equation}

Where:

\begin{align*}
h(\rho)&=-\frac{\left(\rho _1^2+\left(\rho _2+\rho _3\right) \rho _1-\rho _2 \rho _3\right)
  }{\sqrt{\left(\rho
   _2-\rho _1\right) \rho _3}}  F\left(\left.\sin ^{-1}\left(\sqrt{\frac{\rho _1 \left(\rho -\rho _3\right)}{k
   \left(\rho -\rho _1\right) \rho _3}}\right)\right|k\right)+  \\&   -\sqrt{\left(\rho _2-\rho _1\right) \rho _3}
   E\left(\left.\sin ^{-1}\left(\sqrt{\frac{\rho _1 \left(\rho -\rho _3\right)}{k
   \left(\rho -\rho _1\right) \rho _3}}\right)\right|k\right)+ \\&  +\frac{\left(\rho _1-\rho
   _3\right) \left(\rho _1+\rho _2+\rho _3\right) }{\sqrt{\left(\rho
   _2-\rho _1\right) \rho _3}} \Pi \left(\frac{k \rho _3}{\rho
   _1};\left.\sin ^{-1}\left(\sqrt{\frac{\rho _1 \left(\rho -\rho _3\right)}{k
   \left(\rho -\rho _1\right) \rho _3}}\right)\right|k\right)  +  \\& -\sqrt{\frac{\rho  \left(\rho -\rho _2\right) \left(\rho
   _3-\rho \right)}{\rho -\rho _1}}
\end{align*}

and:
\begin{equation}
k=\frac{\rho_1 (\rho_3-\rho_2)}{(\rho_1-\rho_2) \rho_3}
\end{equation}

The negative $k$ value cannot create difficulty being available an analytic continuation of elliptic integrals of first kind also to negative modulus values (see \cite{abr} pag. 593 formulae 17.4.17 and 17.4.18 and pag. 600).

Our first step to solve the radial time problem led us to get time as an addition \eqref{baba} of elliptic integrals of first, second, third kind - i.e. $E, F,\Pi$ - whose modulus $k$ is completely defined by the three real roots of the $\rho$-cubic.

\section{Construction of the inverse function \texorpdfstring{$\rho=\rho(t)$}{TEXT}}
The $t(\rho)$, \eqref{baba} cannot be inverted in any way nor by elementary nor by known special functions; furthermore the Lagrange series reversion theorem would be complicated in that it is defined around a certain point $c$ of the interval in which the function is defined. Typically the point $c = 0$ is used for easier calculations, but this cannot be done in this case since the vector radius $\rho$ magnitude never takes the null value.

Nevertheless, it is quite easy to compute the period of $\rho(t)$ inverse of \eqref{baba} by integrating \eqref{rhointegral} between $\rho=\rho_2$ and $\rho=\rho_3$. 

Defining:
\begin{equation*}
q(\rho)=\frac{1}{\sqrt{\left|E_0\right|}}\frac{\rho^{3/2}}{\sqrt{(\rho-\rho_1)(\rho-\rho_2)(\rho_3-\rho),}}
\end{equation*}
then the a.m. period will be given by:
\begin{equation*}
T=2\int_{\rho_2}^{\rho_3}q(\rho)d\rho=\frac{2}{\sqrt{|E_0|}}[h(\rho_3)-h(\rho_2)]=-\frac{2h(\rho_2)}{\sqrt{|E_0|}}
\end{equation*}
namely:

\begin{align*}
&\frac{\sqrt{|E_0|\rho_3\left(\rho _2-\rho _1\right) }}{2}\cdot T=
 \left[\rho _1^2+\rho _1 \rho _2+\left(\rho _1-\rho _2\right) \rho _3\right]{\bf K}\left(k\right)+\\&+\left(\rho _2-\rho _1\right) \rho_3 E\left(k\right)+\left(\rho
   _3-\rho _1\right) \left(\rho _1+\rho _2+\rho _3\right) \Pi \left(k\rho_3/\rho_1|k\right)
\end{align*}
In order to perform the inversion of \eqref{baba} let us employ the Fourier series, i.e. we are going to construct the Fourier expansion of the function $\rho=\rho(t)$ founding on our knowledge of its period $T$ and of its\footnote{Notice that herefrom and later we use the notation $f^{-1}(t)$ instead of  ${\rm Inv} (f(t))$} inverse $t(\rho)=\rho^{-1}(t)$.

Let us define:
\begin{align*}
t^+(\rho,\rho_0,t_0)&=t_0 + \frac{1}{\sqrt{|E_0|}}[h(\rho)-h(\rho_0)] \\
t^-(\rho,\rho_0,t_0)&=t_0 - \frac{1}{\sqrt{|E_0|}}[h(\rho)-h(\rho_0)]
\end{align*}
If $\rho_0 \in [\rho_2,\rho_3]$  is the starting condition, then during a period $T$ the satellite will perform an ordered motion as $\rho_0 \rightarrow \rho_3\rightarrow \rho_2 \rightarrow \rho_0$; accordingly the instant of time the satellite leaces the major deferent $\rho=\rho_3$ is $\hat{t}_1=t^+(\rho_3,\rho_0,t_0)$and that whenleaves the minor deferent for return at $\rho_0$ will be $\hat{t}_2=t^-(\rho_2,\rho_3,\hat{t}_1)$. Therefore, putting $\zeta_n=2 \pi n /T$ we have: $\hat{a}_n=T a_n/2$ and $\hat{b}_n=T b_n/2$ so that:
\begin{align*}
\hat{a}_0&=\int_0^{T}\rho(t)dt=2\int_{\rho_2}^{\rho_3}s \frac{\partial \rho^{-1}(s)}{\partial s}ds=2\int_{\rho_2}^{\rho_3}s q(s)ds
\end{align*}
where the change of variable has been done $t=\rho^{-1}(s)$ so that $dt=\frac{\partial \rho^{-1}(s)}{\partial s} ds=q(s)ds$. Furthermore we have:
\begin{align*}
\hat{a}_n&=\int_{\rho_0}^{\rho_3}s q(s) \cos \left(\zeta_n  t^+(s,\rho_0,t_0)\right){\rm d}s-\int_{\rho_3}^{\rho_2}s q(s) \cos \left( \zeta_n  t^-(s,\rho_3,\hat{t}_1)\right){\rm d}s+\\&+\int_{\rho_2}^{\rho_0}s q(s) \cos \left( \zeta_n  t^+(s,\rho_2,\hat{t}_2)\right){\rm d}s=
\left[\frac{s}{\zeta_n} \sin \left( \zeta_n  t^-(s,\rho_3,\hat{t}_1)\right) \right]_{\rho_3}^{\rho_2}+\\&
-\frac{1}{\zeta_n}\int_{\rho_3}^{\rho_2}\sin \left( \zeta_n  t^-(s,\rho_3,\hat{t}_1)\right){\rm d}s+
\left[\frac{s}{\zeta_n} \sin \left(\zeta_n  t^+(s,\rho_0,t_0)\right) \right]_{\rho_2}^{\rho_3}+\\
&-\frac{1}{\zeta_n}\int_{\rho_2}^{\rho_3}\sin \left(\zeta_n  t^+(s,\rho_0,t_0)\right){\rm d}s
\end{align*}
But being: $t^-(\rho_2,\rho_3,\hat{t}_1)-t^-(\rho_3,\rho_3,\hat{t}_1)=T/2$:
\begin{align*}
&\left[\frac{s}{\zeta_n} \sin \left( \zeta_n  t^-(s,\rho_3,\hat{t}_1)\right) \right]_{\rho_3}^{\rho_2}=\\&=\frac{\rho_2}{\zeta_n}\sin \left( \zeta_n  t^-(\rho_2,\rho_3,\hat{t}_1)\right)-\frac{\rho_3}{\zeta_n}\sin \left( -n \pi+\zeta_n  t^-(\rho_2,\rho_3,\hat{t}_1)\right)=\\
&=\frac{1}{\zeta_n}\sin \left( \zeta_n  t^-(\rho_2,\rho_3,\hat{t}_1)\right)[\rho_2-(-1)^n \rho_3]
\end{align*}
With $t^+(\rho_3,\rho_0,t_0)-t^{+}(\rho_2,\rho_0,t_0)=T/2$:
\begin{equation*}
\left[\frac{s}{\zeta_n} \sin \left(\zeta_n  t^+(s,\rho_0,t_0)\right) \right]_{\rho_2}^{\rho_3}=\frac{1}{\zeta_n}\sin \left(\zeta_n  t^+(\rho_2,\rho_0,t_0)\right)[(-1)^n \rho_3-\rho_2]
\end{equation*}

Our conclusion is:
\begin{align*}
\hat{a}_n&=\frac{1}{\zeta_n}[\rho_2-(-1)^n \rho_3]\left[\sin \left( \zeta_n  t^-(\rho_2,\rho_3,\hat{t}_1)\right)-\sin \left(\zeta_n  t^+(\rho_2,\rho_0,t_0)\right) \right]+\\
&-\frac{1}{\zeta_n}\left[\int_{\rho_3}^{\rho_2}\sin \left( \zeta_n  t^-(s,\rho_3,\hat{t}_1)\right){\rm d}s+\int_{\rho_2}^{\rho_3}\sin \left(\zeta_n  t^+(s,\rho_0,t_0)\right){\rm d}s \right]=\\
\\&=-\frac{2}{\zeta_n}\cos(\zeta_n \hat{t}_1)\int_{\rho_2}^{\rho_3}\sin\left(\zeta_n \frac{h(s)}{\sqrt{|E_0|}}\right)ds
\end{align*}
An analogous treatment will lead to:
\begin{align*}
\hat{b}_n&=\frac{1}{\zeta_n}[\rho_2-(-1)^n \rho_3]\left[\cos \left(\zeta_n  t^+(\rho_2,\rho_0,t_0)\right)-\cos \left( \zeta_n  t^-(\rho_2,\rho_3,\hat{t}_1)\right) \right]+\\
&-\frac{2}{\zeta_n}\sin(\zeta_n \hat{t}_1)\int_{\rho_2}^{\rho_3}\sin\left(\zeta_n \frac{h(s)}{\sqrt{|E_0|}}\right)ds=-\frac{2}{\zeta_n}\sin(\zeta_n \hat{t}_1)\int_{\rho_2}^{\rho_3}\sin\left(\zeta_n \frac{h(s)}{\sqrt{|E_0|}}\right)ds
\end{align*}

Notice that $\hat{b}_n$ and $\hat{a}_n$ are not independent:
\begin{equation*}
\hat{b}_n=\tan(\zeta_n \hat{t}_1)\hat{a}_n,\quad \hat{a}_n=- \frac{2\cos(\zeta_n \hat{t}_1)}{\zeta_n}\int_{\rho_2}^{\rho_3}\sin\left(\zeta_n \frac{h(s)}{\sqrt{|E_0|}}\right)ds 
\end{equation*}
In such a way the radial time law is be expressed by:
$$
\rho(t)=\frac{2}{T}\left[\frac{\hat{a}_0}{2}+\sum_{n=1}^{+\infty}\left[\hat{a}_n \cos(\zeta_n t)+\hat{b}_n \sin(\zeta_n t)\right] \right]
$$
Then finally:
\begin{equation}
\label{rhotime}
\boxed{\rho(t)=\frac{2}{T}\left[\frac{\hat{a}_0}{2}+\sum_{n=1}^{+\infty}\frac{\hat{a}_n}{\cos(\zeta_n \hat{t}_1)}\cos[\zeta_n(t-\hat{t}_1)] \right]}
\end{equation}

The radial speed $\dot{\rho}(t)$ is provided by the series: 
\begin{equation}
\label{dotrhotime}
\dot{\rho}(t)=-\frac{2}{T}\sum_{n=1}^{+\infty}\frac{\hat{a}_n \zeta_n}{\cos(\zeta_n \hat{t}_1)}\sin[\zeta_n(t-\hat{t}_1)]
\end{equation}

In order to analyze the amplitude of $\dot{\rho}(t)$ , being $\dot{\rho}(\rho)=1/q(\rho)$, putting to zero its derivative with respect to $\rho$, we get
\begin{equation*}
-(\rho_1+\rho_2+\rho_3)\rho^2+2(\rho_1 \rho_2+\rho_1 \rho_3+\rho_2 \rho_3)\rho-3\rho_1\rho_2 \rho_3=0
\end{equation*}
whose only solution $\bar{\rho} \in [\rho_2,\rho_3]$ is:
\begin{equation*}
\bar{\rho}=\frac{\rho _2 \rho _3+\rho _1 \left(\rho _2+\rho _3\right)+\sqrt{\rho
   _1^2 \rho _2^2-\rho _1 \left(\rho _1+\rho _2\right) \rho _3 \rho
   _2+\left(\rho _1^2-\rho _2 \rho _1+\rho _2^2\right) \rho _3^2}}{\rho
   _1+\rho _2+\rho _3}
\end{equation*}

so that:
\begin{equation}
-\frac{1}{q(\bar{\rho})}\leq\dot{\rho}(r)\leq \frac{1}{q(\bar{\rho})},\quad A_{\dot{\rho}}=\frac{1}{q(\bar{\rho})}
\end{equation}
where $A_{\dot{\rho}}$ is the relevant amplitude of $\dot{\rho}$  .

\section{Orbital shapes}
Whilst radius changes during time according to the previous laws, the azimuth will change and in order to detect how it does, we start from the  equation \eqref{areal}; we get:
 \begin{equation} \label {tempusculum}
{\rm d}t=\frac{1}{k_0}\rho^2 {\rm d}\theta 
\end{equation}
Comparing with \eqref{rhointegral} and eliminating time, one obtains the link between the infinitesimal changes of $\theta$ and $\rho$ as a function of $\rho$ only:
\begin{equation}\label{dteta}
\frac{{\rm d}\theta}{{\rm d}\rho}=\pm \frac{k_0}{\sqrt{\left|E_0\right|}} \,\,\frac{1}{\sqrt{\rho(\rho-\rho_1)(\rho-\rho_2)(\rho_3-\rho)}}
\end{equation}
Integrating: 

\begin{equation*}
\theta(\rho)=\theta_0\pm\frac{k_0}{\sqrt{|E_0|}} [f(\rho)-f(\rho_0)]
\label {thetafunction}
\end{equation*}
where:

\begin{equation*}
f(\rho)=-\frac{2}{\sqrt{\rho_3 (\rho_2-\rho_1)}} \cdot F\left(\left.\sin ^{-1}\left(\sqrt{\frac{
  \rho_1 (\rho-\rho_3)}{(\rho-\rho_1)\rho_3 k }}\right)\right| k\right)
\end{equation*}

which can be inverted as:

\begin{equation} \label{rhotheta}
\boxed{\rho(\theta)=\rho_3 \cdot \frac{1-k\cdot \text{sn}^2\left(g(\theta),k
\right)}{1-(\rho_3/\rho_1) k\cdot\text{sn}^2\left(g(\theta),k\right)}}
\end{equation}

with:
\begin{equation*}
g(\theta)=\frac{(\sqrt{|E_0|}/k_0)(\theta-\theta_0)+f(\rho_0)}{2} \sqrt{\rho_3(\rho_2-\rho_1)}
\end{equation*}
The function $\rho=\rho(\theta)$ is periodic with an angular period given by:
\begin{align*}
\hat{T}=2k_0\int_{\rho_2}^{\rho_3}\frac{q(\rho)}{\rho^2}{\rm d}\rho=-\frac{2k_0 f(\rho_2)}{\sqrt{|E_0|}}=\frac{4k_0 K(k)}{\sqrt{|E_0|\rho_3(\rho_2-\rho_1)}}
\end{align*}
Therefore $\hat{T}\neq 2\pi$ so that the mobile trajectory will not be closed-as in the keplerian case- because at each turn (i.e. when $\theta$ is added of $2\pi$), $\rho$ does not take again the same value: then there is a displacement which is building up during the revolutions and this will produce the orbital drifting.

\section{A further inversion: azimuth-time law \texorpdfstring{$\theta=\theta(t)$}{TEXT}}
With reference to the approach given above, let us compute the Fourier expansion of $\theta(t)$; defining $\hat{\alpha}_n=T\alpha_n/2$ e $\hat{\beta}_n=\tan(\zeta_n \hat{t}_1)\hat{\alpha}_n$ we get:
\begin{align*}
\hat{\alpha}_0&=2\int_{\rho_2}^{\rho_3}\frac{q(s)}{s^2}{\rm d}s=\frac{4K(k)}{\sqrt{|E_0|\rho_3(\rho_2-\rho_1)}}\\ \hat{\alpha}_n &=	\frac{4\cos(\zeta_n \hat{t}_1)}{\zeta_n}\int_{\rho_2}^{\rho_3}\frac{1}{s^3}\sin\left(\zeta_n \frac{h(s)}{\sqrt{|E_0|}}\right){\rm d}s
\end{align*}
By means of such coefficients we can expand in Fourier series the function $1/\rho^2(t)$, and multiplying by $k_0$ and integrating, we get $\theta(t)$: 

$$
\theta(t)-\theta_0=\frac{2k_0}{T}\left[\frac{\hat{\alpha}_0}{2}\psi+\sum_{n=1}^{+\infty}\frac{\hat{\alpha}_n}{\zeta_n \cos(\zeta_n \hat{t}_1)}\sin[\zeta_n(\psi-\hat{t}_1)] \right]_{\psi=t_0}^{\psi=t},
$$
or:
\begin{equation}\label{thetati}
\boxed{\theta(t)-\theta_0
=\frac{2k_0}{T}\left[\frac{\hat{\alpha}_0}{2}(t-t_0)+\sum_{n=1}^{+\infty} \frac{\hat{\alpha}_n}{\zeta_n }\cdot\frac{\sin[\zeta_n(t-\hat{t}_1)]-\sin[\zeta_n(t_0-\hat{t}_1)]}{\cos(\zeta_n \hat{t}_1)}   \right]}
\end{equation}

Angular speed and acceleration do change in time during satellite revolutions according to:
\begin{equation}\label{dotthetati}
\ddot{\theta}(t)=-\frac{2k_0}{T}\sum_{n=1}^{+\infty} \hat{\alpha}_n\zeta_n \cdot\frac{\sin[\zeta_n(t-\hat{t}_1)]}{\cos(\zeta_n \hat{t}_1)}  
\end{equation}
Remembering $\zeta_n=2 \pi n/T$ we have $\ddot{\theta}(t)=0$ for $t=\hat{t}_1$ and $t=\hat{t}_1+T/2$ corresponding to a minimum and to a maximum of $\dot{\theta}(t)$ respectively:
\begin{equation*}
\frac{2k_0}{T}\left[\frac{\hat{\alpha}_0}{2}+\sum_{n=1}^{+\infty}\frac{\hat{\alpha}_n}{\cos(\zeta_n \hat{t}_1)}   \right]\leq \dot{\theta}(t) \leq\frac{2k_0}{T}\left[\frac{\hat{\alpha}_0}{2}+\sum_{n=1}^{+\infty}\frac{(-1)^n\hat{\alpha}_n}{\cos(\zeta_n \hat{t}_1)}   \right]
\end{equation*}
The signal amplitude is then:
\begin{equation}
2A_{\dot{\theta}}=-\frac{4k_0}{T}\sum_{n=0}^{+\infty}\frac{\hat{\alpha}_{2n+1}}{\cos(\zeta_{2n+1} \hat{t}_1)} 
\end{equation}

As it concerns the tangential velocity $v_\theta$ defining $\hat{\gamma}_n=T\gamma_n/2$ we get:
\begin{align*}
\hat{\gamma}_0&=2\int_{\rho_2}^{\rho_3}\frac{q(s)}{s}{\rm d}s=4\cdot\frac{\rho_1 K(k)+(\rho_3-\rho_1)\Pi\left( k \rho_3/\rho_1|k\right)}{\sqrt{|E_0|\rho_3(\rho_2-\rho_1)}}\\ \quad \hat{\gamma}_n&=\frac{2\cos(\zeta_n \hat{t}_1)}{\zeta_n}\int_{\rho_2}^{\rho_3}\frac{1}{s^2}\sin\left(\zeta_n \frac{h(s)}{\sqrt{|E_0|}}\right){\rm d}s
\end{align*}
Therefrom:
\begin{equation}\label{vthetati}
v_{\theta}(t)=\frac{2k_0}{T}\left[\frac{\hat{\gamma}_0}{2}+\sum_{n=1}^{+\infty}\frac{\hat{\gamma}_n}{\cos(\zeta_n \hat{t}_1)}\cos[\zeta_n(t-\hat{t}_1)] \right]
\end{equation}
As usually we have:
\begin{equation*}
\frac{2k_0}{T}\left[\frac{\hat{\gamma}_0}{2}+\sum_{n=1}^{+\infty}\frac{\hat{\gamma}_n}{\cos(\zeta_n \hat{t}_1)}   \right]\leq v_{\theta}(t) \leq\frac{2k_0}{T}\left[\frac{\hat{\gamma}_0}{2}+\sum_{n=1}^{+\infty}\frac{(-1)^n\hat{\gamma}_n}{\cos(\zeta_n \hat{t}_1)}   \right]
\end{equation*}
Therefore this signal amplitude will be given by:
\begin{equation*}
2A_{v_\theta}=-\frac{4k_0}{T}\sum_{n=0}^{+\infty}\frac{\hat{\gamma}_{2n+1}}{\cos(\zeta_{2n+1} \hat{t}_1)} 
\end{equation*}

\section{The keplerian problem as a benchmark of our inversion method}
In this section we are giving a theoretical contribution about the method explained above: its validation by the classic keplerian benchmark.

In radial sense the motion equation is:
\begin{equation*}
\frac{{\rm d}}{{\rm d}t}\left(\frac{\dot{\rho}^2}{2}+\frac{k_0^2}{2\rho^2}-\frac{\mu}{\rho} \right)=0
\end{equation*}
We have:
\begin{equation*}
\epsilon_0=\frac{\dot{\rho}_0^2}{2}+\frac{k_0^2}{2\rho_0^2}-\frac{\mu}{\rho_0}
\end{equation*}
Therefore:
\begin{equation*}
\frac{\dot{\rho}^2}{2}+\frac{k_0^2}{2\rho^2}-\frac{\mu}{\rho}=-|\epsilon_0|
\end{equation*}
Therefore:
\begin{equation*}
{\rm d}t=\pm \frac{\rho {\rm d}\rho}{\sqrt{-2|\epsilon_0|\rho^2+2 \mu \rho-k_0^2}}
\end{equation*}
which can be written as:
\begin{equation*}
{\rm d}t=\pm\frac{1}{\sqrt{2|\epsilon_0|}}\frac{\rho {\rm d}\rho}{\sqrt{(\rho-\rho_1)(\rho_2-\rho)}}
\end{equation*}
So that:
\begin{equation*}
t-t_0=\pm\frac{1}{\sqrt{2|\epsilon_0|}}[H(\rho)-H(\rho_0)]
\end{equation*}
with:
\begin{align*}
H(\rho)&=-\sqrt{\left(\rho -\rho _1\right) \left(\rho _2-\rho
   \right)}-\frac{1}{2} \left(\rho _1+\rho _2\right) \tan
   ^{-1}\left(\frac{-2 \rho +\rho _1+\rho _2}{2 \sqrt{\left(\rho -\rho
   _1\right) \left(\rho _2-\rho \right)}}\right)
\end{align*}
Function $\rho(t)$ will have a period given by:
\begin{equation*}
T=\frac{2}{\sqrt{2|\epsilon_0|}}[\lim_{\rho\rightarrow \rho_2}H(\rho)-\lim_{\rho\rightarrow \rho_ 1}H(\rho)]=\frac{\pi(\rho_1+\rho_2)}{\sqrt{2|\epsilon_0|}}
\end{equation*}
The instant of time when the satellite leaves the major deferent $\rho_2$ towards the minor, i.e. $\rho_1$ is:
\begin{equation*}
\hat{t}_1=t_0+\frac{T}{4}-\frac{H(\rho_0)}{\sqrt{2|\epsilon_0|}}
\end{equation*}
As previously we have:
\begin{align*}
\hat{a}_0=\frac{\pi(3 \rho _1^2+2 \rho _2 \rho _1+3 \rho _2^2)}{4\sqrt{2|\epsilon_0|}}
\end{align*}
So that:
\begin{equation*}
\hat{a}_n=-\frac{2\cos(\zeta_n \hat{t}_1)}{\zeta_n} \int_{\rho_1}^{\rho_2}\sin\left[\zeta_n\left(\frac{H(s)-T/4}{\sqrt{2|\epsilon_0|}} \right) \right]{\rm d}s, \quad \hat{b}_n=\tan(\zeta_n \hat{t}_1)\hat{a}_n
\end{equation*}
The solution can be expanded in Fourier series and referring to the previous notation we have:
\begin{equation*}
\hat{\alpha}_0=\frac{2\pi}{\sqrt{2\rho_1\rho_2|\epsilon_0|}},\quad \hat{\alpha}_n=\frac{4\cos(\zeta_n \hat{t}_1)}{\zeta_n} \int_{\rho_1}^{\rho_2}\frac{1}{s^3}\sin\left[\zeta_n\left(\frac{H(s)-T/4}{\sqrt{2|\epsilon_0|}} \right) \right]{\rm d}s
\end{equation*}
The roots of the quadratic equation obtained putting  $-2|\epsilon_0|\rho^2+2 \mu \rho-k_0^2=0$ are:
\begin{equation*}
\rho_{1,2}=\frac{\mu \pm \sqrt{\mu^2-2|\epsilon_0|k_0^2}}{2| \epsilon_0|},
\end{equation*}
so that:
\begin{equation*}
\rho_1\cdot \rho_2=\frac{k_0^2}{2|\epsilon_0|},\quad \rho_1+\rho_2=\frac{\mu}{|\epsilon_0|},\quad \rho_1^2+\rho_2^2=\frac{\mu^2-|\epsilon_0| k_0^2}{|\epsilon_0|^2}
\end{equation*}
and summarizing:
\begin{equation*}
\hat{a}_0=\frac{\pi(3\mu^2-2|\epsilon_0|k_0^2)}{(2|\epsilon_0|)^{5/2}}, \quad \hat{\alpha}_0=\frac{2\pi}{k_0}
\end{equation*}
Finally:
\begin{equation*}
\hat{\gamma}_0=\frac{2\pi}{\sqrt{2|\epsilon_0|}},\quad \hat{\gamma}_n=\frac{2\cos(\zeta_n \hat{t}_1)}{\zeta_n} \int_{\rho_1}^{\rho_2}\frac{1}{s^2}\sin\left[\zeta_n\left(\frac{H(s)-T/4}{\sqrt{2|\epsilon_0|}} \right) \right]{\rm d}s
\end{equation*}
By above we are allowed to obtain the time laws $\dot{\rho}(t)$, $\theta(t)$, $\dot{\theta}(t)$ e $v_{\theta}(t)$.
The trajectory's polar equation will be given by:
\begin{align*}
\theta-\theta_0&=\frac{k_0}{\sqrt{2|\epsilon_0|}}\int_{\rho_0}^{\rho} \frac{{\rm d}s}{s\sqrt{(s-\rho_1)(\rho_2-s)}}=\\&=\frac{2k_0}{\sqrt{2|\epsilon_0|\rho_1\rho_2}}\left[ \sin ^{-1}\left(\sqrt{\frac{\rho_2 (\rho_1-s)}{s(\rho_1-\rho_2)}}\right) \right]_{\rho_0}^{\rho}=2\left[ \sin ^{-1}\left(\sqrt{\frac{\rho_2 (\rho_1-s)}{s(\rho_1-\rho_2)}}\right) \right]_{\rho_0}^{\rho}
\end{align*}
The period of function $\rho(\theta)$ is:
\begin{equation*}
\hat{T}=\frac{2k_0}{\sqrt{2|\epsilon_0|}}\int_{\rho_1}^{\rho_2} \frac{{\rm d}s}{s\sqrt{(s-\rho_1)(\rho_2-s)}}=2\pi
\end{equation*}
Putting:
\begin{equation*}
\hat{g}(\theta)=\theta-\theta_0+2 \sin ^{-1}\left(\sqrt{\frac{\rho_2 (\rho_1-\rho_0)}{\rho_0(\rho_1-\rho_2)}}\right)
\end{equation*}
We get:
\begin{align*}
\rho(\theta)&=\frac{2 \rho_1 \rho_2}{\cos (\hat{g}(\theta)) (\rho_2-\rho_1)+\rho_1+\rho_2}=\\
&=\frac{\dfrac{2\rho_1\rho_2}{\rho_1+\rho_2}}{1+\dfrac{\rho_2-\rho_1}{\rho_1+\rho_2}\cos(\hat{g}(\theta))}=\frac{k_0^2/\mu}{1+e \cos(\hat{g}(\theta))}
\end{align*}
which is the well-known polar law of a keplerian orbit but generalized to the case of a starting point at the perigee ($\theta_0=0$ e $\rho_1=\rho_0$)  and where:
\begin{equation*}
e=\dfrac{\rho_2-\rho_1}{\rho_1+\rho_2}
\end{equation*}
is the ellipse eccentricity.

\section{Sample problem and output analysis}
In this section we will apply the main formulae found above for the $J_2$ problem to a practical case. First of all we consider a satellite in its unperturbed motion along a keplerian orbital ellipse; such a status is the startup condition. At a certain instant, which is our time origin, we imagine the bulge potential will act perturbing the previous dynamics.

 Accordingly, let the satellite move along a keplerian ellipse of eccentricity $e=0.3$ on the Earth equatorial plane with a specific angular momentum of $k_0=95000$ $\text{Km}^2/\text{s}$. Start at $t_0=0$ $\text{s}$ with $\theta_0=40$ $\deg$, we have:
\begin{align*}
\rho_0&=\frac{k_0^2/\mu}{1+e\cos(\theta_0)}\approx 18410.7 \quad \text{Km}\\ \dot{\rho}_0&=\frac{\mu}{k_0}e \sin(\theta_0)\approx 0.8091 \quad \text{Km}/\text{s} \\ 
   \dot{\theta}_0&=\frac{\mu}{k_0\rho_0}(1+e\cos(\theta_0))\approx 0.0161 \quad \deg/\text{s}
   \end{align*} 
  Let it be:
\begin{equation*}
P_3(\rho)=-\rho^3+\frac{2\mu}{\left|E_0\right|}\rho^2-\frac{k_0^2}{\left|E_0\right|}\rho+\frac{2\mu\eta}{\left |E_0\right|}
\end{equation*}
Accordingly $|E_0|\approx 16.023$ $Km^2/s^2$ and solving $P_3(\rho)=0$ we found the roots:
\begin{equation*}
\rho_1\approx 1.94542 \quad \text{Km}, \quad \rho_2\approx 17416.1 \quad \text{Km}, \quad \rho_3\approx 32335.3 \quad \text{Km}
\end{equation*}
General bounded motion has then both lower and upper bounds so that the particle cannot approach near than some minimum or move farther than some maximum distance; furthermore the angular velocity has a constant sign (see fig.6), same as that of the moment of momentum throughout the motion. As anticipated, the first root
falls within the Earth. Following our notations, we forecast a bounded trajectory confined in an annulus between $\rho_2$ e $\rho_3$ with two time laws $\theta(t)\in [0,2\pi]$ and $\rho(t)\in[\rho_2,\rho_3]$ with period $T\approx 39048.1$  $s$. 
The time scanned is approximately 2.4 days corresponding to slightly less than 5 turns.

The main outputs of this paper are the analytic ones in framed formulae.We also performed analytical computations of them obtaining some interesting plots listed below. They have been omitted here,but the interested readers can in any case ask the mailing author for them and for the relevant program scripts in Matlab. So they will appreciate how much our analytical outputs comply with the numerical computations which are within everyne's grasp.

\begin{enumerate}
\item {\bf Radial distance.}
The radial distance $\rho (t)$ of the satellite from the Earth oscillates between the boundaries  $\rho_{1}$ and $\rho_{2}$ indefinitely, according to  \eqref{rhotime}

\item {\bf Radial speed.}
Of course the $\dot\rho$ sign changes for being the orbit not a crcle and then formula \eqref{dotrhotime} forecasts  alternate signs for time derivative of distance particle - Earth.
The derivative's null values refer to those satellite times when the particle touches the deferent circles.

\item {\bf Azimuth versus time.}
The azimuth is computed, see \eqref{thetati}, by means of inverse functions and then it starts again from zero at the end of each turn of 0.48 days.

\item {\bf Tangential speed versus time.}
This behavior is coming from formula\eqref{vthetati}

\item {\bf Angular velocity.}
As told before, the sense of orbital trajectory is always the same so that the time derivative of azimuth is always greater than zero.

\item {\bf Trajectory.}
We saw a typical bounded set of orbits with $\theta$ advancing as $\rho$ oscillates between inner and upper circle, according to \eqref{rhotheta}.
In the classic-no bulge-non relativistic Kepler problem, a particle eternally follows the same perfect ellipse. The presence of the bulge acts as if a third force attracting the particle were added to keplerian gravitation, especially for small radii. This third force causes the quasi-elliptical orbit of the particle to be not fixed but undergoes a precession  in the direction of its rotation, so forming a small rose. This effect was also measured in the planets Mercury, Venus and Earth under the attraction of the Sun mass which is flattened too. 
The same precession effect is found in the mathematical model of General Relativity in which the bulge is not taken into account at all, but only the space-time curvature and the orbit comes as geodesic path within it. 
\end{enumerate}

\section{Conclusions}
The report analyzes the possibility of inverting and solving the motion equations when the Earth's oblateness rules a satellite's orbit. During our treatment several assumptions have been sparsely done and collected below:
\begin{enumerate}
\item Attractor constant density.
\item Axisymmetric spheroidal attraction, i. e.  not dependent on azimuth.
\item Satellite not too close to the planet: $\rho>>R$.
\item Potential higher order terms beyond $J_2$ really negligible.
\item Effects of atmospheric drag ignored. 
\item Effects of Sun radiation pressure ignored. 
\item Satellite motion bounded (total negative energy).
\item Neglected any gravitational effect by the satellite on the spheroid and those of any other celestial body (Sun, Moon) on the satellite.
\item The spherical frame of reference origin is taken at the centre of spheroidal mass whose only motion is therefore just the rotation -around its North-South axis- responsible of the flatness.
\item Each computation can be carried out knowing at all 7 quantities, namely:
$\rho_0,\dot\rho_0,\theta_0,\omega_0$, as it concerns the satellite startup
and: $\mu, R, J_2$, as it concerns the spheroid.
\end{enumerate}
On such assumptions, the following conclusions have been met.
New formulae have been obtained for solutions describing the "planarized" motion of a particle attracted in vacuo by an oblate gravitating spheroid. Our solutions do not stem from a perturbative approach but have been carried out facing the problem as a whole in its nonlinear nature and then leading to elliptic integrals of I, II, III kind. They therefore cannot be inverted by means of any special functions: nevertheless such inverse have been given by constructing their period and their Fourier expansion: then the planarized $J_2$ problem is fully solved in the sense of Liouville having we obtained the Fourier series expansion of the inversions.

Our approach has been successfully validated by applying it to the basic  keplerian case as a benchmark.
A sample problem has been finally presented where some outputs have been plotted and discussed in their physical meaning.

According to General Relativity, following the Schwarzschild model of a M-spherical non rotating attractor, the particle motion analysis in a $(\rho,\theta)$ frame leads to a dynamical forcing term given by:
$$
\mu\left(\dfrac{1}{\rho^2}+3\dfrac{k^2}{c^2\rho^4}\right)
$$
per unity attracted mass, instead of \eqref{syst}:
$$
\mu\left(\dfrac{1}{\rho^2}+3\dfrac{\eta}{\rho^4}\right)
$$
Therefore the oblateness -modeled in $J_2$ fashion- drives in a classical context to an ODE system having the same mathematical structure of the Schwarzschild planetary motion solution (1916) which allowed to explain the 43''/century precession which had been outside the possibilities of any pre-relativistic calculation.
Of course our inversions can easy be used -after little adjustment-in order to perform some practical evaluation cases in Schwarzschild context, free from numerical methods and selecting how many terms to include in expansions for $\rho(t) $ and $\theta(t)$.


\section{acknowledgements}

The authors wish to thank D. Ritelli who inspired a valuable choice at the beginning of this work.

\section*{Conflict of interest}
The authors declare that they have no conflict of interest.

\printbibliography

@article{hoo,
author={Hoots,F.R. and France, R.G.},
title={An analytic satellite theory using gravity and a dynamic atmosphere},
journal={Celestial Mechanics},
year={1987},
volume={--},
number={40},
pages={1-18},
}

@article{s,
  title={Hypergeometric {F}unctions of {T}hree {V}ariables},
  author={Saran, Shanti },
  journal={Ganita},
  number={1}, 
  volume={5},
  pages={77--91},
  year={1954}
}

@article{Minrit,
  title={Motions about a fixed point by hypergeometric functions: new non-complex analytical solutions and integration of herpolhode.},
  author={Mingari Scarpello,G. and Ritelli, D.},
  journal={Celestial Mechanics and Dynamical Astronomy},
  number={130}, 
  volume={42},
  pages={1-34},
  year={2018}
  }

@article{gurfil,
   title={Analytical solutions for ${J_2}$-perturbed unbounded equatorial orbits},
  author={Martinusi, Vladimir and Gurfil, Pini},
  journal={Celestial Mechanics and Dynamical Astronomy},
  volume={115},
  number={1},
  pages={35--57},
  year={2013},
  publisher={Springer}
  }

@article{allman,
  title={On the attraction of ellipsoids, with a new demonstration of {C}lairaut's theorem being an account of the late professor {M}ac{C}ullagh's lectures on those subjects},
  author={Allman, G. J. },
  journal={Royal Irish Academy},
  volume={22},
  number={--},
  pages={379-395},
  year={1853},
  }

@article{Anne,
 title={Verified solutions for the gravitational attraction to an oblate spheroid: {I}mplications for planet mass and satellite orbits},
  author={Hofmeister, Anne M. and Criss, Robert E. and Criss, Everett M.},
  journal={Planetary and Space Science},
  volume={152},
  pages={68--81},
  year={2018},
  publisher={Elsevier}
}

@article{Mac,
  title={Periodic orbits about an oblate spheroid},
  author={MacMillan, William Duncan},
  journal={Transactions of the American Mathematical Society},
  volume={11},
  number={1},
  pages={55--120},
  year={1910},
  publisher={JSTOR}
}

@article{cong,
  title={Satellite motion around an oblate planet: a perturbation solution for all orbital parameters},
  author={Danielson, D.A. and Latta, G.E. and Sagovac, C.P. and Krambek, S.D. and Snider, G.R.}, 
   journal={Astrodynamics Conference},
  volume={--},
  number={--},
  pages={1-21},
  year={1990},
  publisher={AIAA}
}

@article{pelag,
  title={Planetary equations based upon {N}ewton's equations of motions and {N}ewton's gravitational field of a static homogeneous oblate spheroidal {S}un},
  author={Omonile, J. and Koffa, D. and Howusu, S.},
   journal={Advances in applied Science Research},
  volume={5},
  number={1},
  pages={282-287},
  year={2014},
  publisher={Pelagia Research Library}
}

@article{jeze,
 title={An analytic solution for the ${J_2}$ perturbed equatorial orbit},
  author={Jezewski, D.J.},
  journal={Celestial mechanics},
  volume={30},
  number={4},
  pages={363--371},
  year={1983},
  publisher={Springer}
}

@article{vinti1,
  title={Representation of the Earth's gravitational potential},
  author={Vinti, John P.},
   journal={Celestial Mechanics},
  volume={4},
  number={3},
  pages={348-367},
  year={1971},
  publisher={Kluwer}
}

@article{celletti,
  title={Non-integrability of the problem of motion around an oblate planet},
  author={Celletti, Alessandra and Negrini, Piero},
  journal={Celestial Mechanics and Dynamical Astronomy},
  volume={61},
  number={3},
  pages={253--260},
  year={1995},
  publisher={Springer}
}

@article{lerman,
  title={Hamiltonian systems with loops of a separatrix of a saddle center},
  author={Lerman, L.M.},
  journal={Selecta math.sov.},
  volume={10},
  number={297},
  %pages={--},
  year={1991},
  publisher={--}
}

@article{sim2,
  title={Measuring the lack of integrability of the ${J}_2$ problem for Earth's satellite},
  author={Simo,C.},
  journal={Predictability, stability and chaos in N-body dynamical systems},
  volume={--},
  number={--},
  %pages={--},
  year={1991},
  publisher={Plenum press}
}

@article{sim,
  title={Non-integrability of the ${J}_2$ problem},
  author={Irigoyen, M. and Simo,C.},
  journal={Celestial Mechanics and Dynamical Astronomy},
  volume={55},
  number={3},
  pages={281},
  year={1993},
  publisher={Springer}

}

@article{lio,
  title={Note sur lÕint\'{e}gration des equations diff\'{e}rentielles de la dynamique},
  author={Liouville, J.},
  journal={Journal de Math\'{e}matiques (Journal de Liouville) },
  volume={XX},
  number={2},
  pages={137-145},
  year={1855},
  publisher={--}
  
}

@article{jans,
 title={Spheroidal {P}otentials and {G}ravitational {A}ttraction by a {R}od and a {D}isc},
  author={Janssens, F.L.},
  journal={Advances in the Astronautical Sciences},
  volume={116},
  pages={1--19},
  year={2004}
}

@article{brouwer,
 title={The motion of a particle with negligible mass under the gravitational attraction of a spheroid},
  author={Brouwer, Dirk},
  journal={The Astronomical Journal},
  volume={51},
  pages={223},
  year={1946},
  publisher={American Astronomical Society}
}

@article{king,
  title={The effect of the {E}arth's oblateness on the orbit of a near satellite},
  author={King-Hele, Desmond G.},
  journal={Proceedings of the Royal Society of London. Series A. Mathematical and Physical Sciences},
  volume={247},
  number={1248},
  pages={49--72},
  year={1958},
  publisher={The Royal Society London}
}

@book{camb,
  title={Introduction to classical integrable systems},
  author={Babelon, O. and Bernard, D. and Talon, M.},
  year={2003},
  publisher={Cambridge University Press},
  address={Cambridge}
  }

@book{puca,
  title={Theory of orbits, vol. II: Perturbative and geometrical methods},
  author={Boccaletti,D. and Pucacco, G.},
  year={1999},
  publisher={Springer},
  address={Berlin}
  }

@book{tod,
  title={A history of mathematical theories   of attraction and figure of the Earth},
  author={Todhunter, I.},
  year={1962},
  publisher={Dover reprint},
  address={New York}
  }

@book{oliver,
  title={Satellite orbits. Models, methods and applications},
  author={Montenbruck, O. and Gill, E.},
  year={2012},
  publisher={Springer},
  address={New York}
}

@book{gold,
  title={Classical {M}echanics},
  author={Goldstein, Herbert and Poole, Charles and Safko, John},
  year={2002},
  publisher={Addison Wesley},
  address={Boston}
}

@book{whit,
  title={A treatise on the analytical dynamics of particles and rigid bodies},
  author={Whittaker, Edmund Taylor},
  year={1944},
  publisher={Dover},
  address={New York}
  
}

@book{teodorescu1,
  title={Particle mechanics, {I}},
  author={Teodorescu, Petre},
  year={2009},
  publisher={Springer},
  address={New-York}
}

@book{abr,
  title={Handbook of mathematical functions with formulas, graphs, and mathematical tables},
  author={Abramowitz, Milton and Stegun, Irene A},
  year={1972},
  publisher={Dover}
}

@article{marzari2020ring,
  title={Ring dynamics around an oblate body with an inclined satellite: the case of Haumea},
  author={Marzari, Francesco},
  journal={Astronomy \& Astrophysics},
  volume={643},
  pages={A67},
  year={2020},
  publisher={EDP Sciences}
}

\end{document}